\documentclass[12pt,a4paper,final]{iopart}

\usepackage{iopams}  
\usepackage{graphicx}
\usepackage[breaklinks=true,colorlinks=true,linkcolor=blue,urlcolor=blue,citecolor=blue]{hyperref}
\usepackage{caption}
\usepackage{subcaption}

\begin{document}

\title{Unitary evolution for a quantum Kantowski-Sachs cosmology}

\author{Sridip Pal$^{1,2}$}
\address{$^1$  Department of Physical Sciences, Indian Institute of Science Education and Research Kolkata, Mohanpur, West Bengal-741246, India}
\address{$^2$ Department of Physics, University of California, San Diego, 9500 Gilman Drive, La Jolla, CA-92093, U.S.A}
\ead{sridippaliiser@gmail.com}

\author{Narayan Banerjee}
\address{ Department of Physical Sciences, Indian Institute of Science Education and Research Kolkata, Mohanpur, West Bengal-741246, India}
\ead{narayan@iiserkol.ac.in}


\begin{abstract}
It is shown that like Bianchi I, V and IX models, a Kantowski-Sachs cosmological model also allows a unitary evolution on quantization. It has also been shown that this unitarity is not at the expense of the anisotropy. Non-unitarity, if there is any, cannot escape notice in this as the evolution is studied against a properly oriented time parameter fixed by the evolution of the fluid. Furthermore, we have constructed a wave-packet by superposing different energy eigenstates, thereby establishing unitarity in a non-trivial way, which is a stronger result than an energy eigenstate trivially  giving time independent probability density. For $\alpha\neq1$, we have proved that the Hamiltonian upon suitable operator ordering admits a self-adjoint extension under reasonable assumption, using a standard theorem.  This consolidates the recently shown fact that the problem of non-unitarity cannot be a generic problem of anisotropy.
\end{abstract}

\vspace{2pc}
\section{Introduction}
The Universe as a whole, being a viable physical system, is expected to have an underlying quantum theory. The interaction that governs the Universe at large scale is gravity, which as yet does not have any consistent and universally accepted quantum theory and thus one cannot really look at the quantum universe as an application of quantum gravity. The alternative, an attempt to bypass the conceptual difficulties of quantum gravity, is provided by Wheeler-deWitt equation\cite{dewitt, wheeler, misner}, which is actually the starting point for quantizing a cosmological model, and forms the subject of quantum cosmology. However, quantum cosmology by itself is not actually free from conceptual problems as well as some practical difficulties. One raging problem is that of the choice of a properly oriented time parameter, as in a relativistic theory time is a coordinate and not a scalar parameter same for all observers. This problem, and the possible resolutions in various ways constitute 
almost a branch of quantum cosmology\cite{kuchar1, isham, rovelli, anderson}. Recently, a choice of an oriented time parameter depending on the evolution of a perfect fluid constituting the matter distribution in the universe has been in use with a fair amount of success. The strategy actually had been given long back by Lapchinskii and Rubakov\cite{rubakov} based on Schutz's formalism of writing the velocity vector of the fluid in terms of some thermodynamic potentials\cite{schutz1, schutz2}. Another serious conceptual problem, that of the interpretation of the wave function and that of the boundary conditions are reviewed by Wiltshire\cite{wilt}, Halliwell\cite{halli} and very recently by Pinto-Neto and Fabris\cite{nelson1}. \\

One nagging problem of quantum cosmology had been the fact that anisotropic models are found to be nonunitary and hence do not respect the conservation of probability. As the collection of all the 3-space Riemannian metric, the Superspace, is too big to work in, the usual practice is to restrict the description in a truncated version of that, namely a minisuperspace. The natural choice of the minisuperspace is indeed a spatially homogeneous and isotropic FRW 3-space which is compatible with the presently observed universe. But anisotropy has its relevance. The formation of the present structure of the Universe requires a small but finite anisotropy in the CMB temperature ($\frac{\Delta T}{T} \sim 10^{-5}$) and the recent observations also reveal the same order of magnitude of anisotropy. The other requirement is aesthetic; there is no reason, a priori, for the universe to be isotropic. So a correct version of a quantum theory should not lead to inconsistent physics, such as a
  nonconservation of probability, even for the anisotropic models. It is interesting to note that without the existence of a time parameter with the correct orientation, this nonconservation of probability in quantum anisotropic models is generally obscure and may escape being detected\cite{lidsey, nelson2}. This alleged pathology of non-unitarity often ascribed to the hyperbolicity of the Hamiltonian of the anisotropic cosmologies\cite{alvarenga3}.\\

Very recently it has been shown that the nagging problem of  non-unitarity can actually be cured. For a Bianchi type I model, either by a clever ordering of operators or by a suitable transformation of coordinates, one can find self-adjoint extension of the Hamiltonian\cite{sridip1}. In fact this is not exclusive for a Bianchi I model where the spatial curvature is zero, the similar self-adjoint extension is possible for anisotropic spaces with negative and positive spatial curvature  given by Bianchi V and IX models as well\cite{sridip2}. It deserves mention that both these investigations involve a properly oriented time parameter through the evolution of the fluid as mentioned earlier. Of course the construction of the wave packet and the evaluation of the norm depends on the degree of difficulty of the integration involved, so one cannot work out what may be called the ``general'' situation, one actually has to depend on some specific equations of state for the fluid present. But one counter example is good enough to disprove the fact that alleged non-unitarity is a generic problem of anisotropic models, and one now has several nontrivial examples. It deserves mention that even istropic models actually can show a nonunitary evolution unless the operator oredring is carefully chosen. Issues of unitarity of a quantized isotropic model in the presence of a scalar field has been very recently discussed by Almeida {\it etal}\cite{almeida}.\\

The motivation of the present work is to check if unitary evolution is a possibility in a Kantowski-Sachs (KS) cosmology, which is qualitatively different from Bianchi I, V and IX models. The KS metric represents a homogeneous but anisotropic spacetime and has its own relevance in gravity in various ways, such as the fact that a KS metric is isometric to the metric for the interior of a spherically symmetric black hole. For a concise list of the relevance of a KS metric, we refer to the recent work by Parisi, Radicella and Vilasi\cite{parisi}. A quantization scheme for a KS metric, although not much talked about, is not completely new. It has been discussed in the context of a non-commutative geometry by Garcia-Compean, Obregon and Ramirez\cite{ramirez}, and very recently in the context of a loop quantum cosmology by Joe and Singh\cite{param}. Quantizing a KS cosmology in a more standard version of gravity was discussed long back by Conradi\cite{conradi}. The latter includes 
 a fluid, namely a pressure-less fluid in the quantization scheme, but does not indicate anything regarding unitarity. \\

We show that the Hamiltonian is indeed self-adjoint in the case of a Kantowski-Sachs model with a stiff fluid (i.e. fluid with an equation of state $P$= $\rho$, where $P$ is pressure and $\rho$ is density of fluid), upon choosing a suitable weight factor while defining the norm. An explicit example of a unitary solution has also been obtained. The complete calculations could be possible, however, only for a particular equation of state, that of a stiff fluid. For other equations of state, we can not find any explicit solution owing to the computational difficulty. Nonetheless, we could show that even for such cases ($P\neq \rho$)the Hamiltonian admits a self-adjoint extension under reasonable assumption with a clever ordering of operators. \\

The general formalism and the Wheeler deWitt equation for the KS model is given in section 2. Section 3 takes up the case of the particular example with a stiff fluid. In section 4, we arrive at the relevant Wheeler deWitt equation for the general barotropic fluid and show how a suitable operator ordering can make Hamiltonian self-adjoint extendible. Section 5 includes some discussions.

\section{Quantization of a Kantowski-Sachs cosmological model}

We start with the standard Einstein-Hilbert action
\begin{equation}\label{Action}
{\mathcal A}=\int_{M}d^{4}x \sqrt{-g}R +2 \int_{\partial M} \sqrt{h}h_{ab}K^{ab}+\int_{M} d^{4}x \sqrt{-g}P,
\end{equation}
where $K^{ab}$ is the extrinsic curvature, $h^{ab}$ the induced metric over the boundary $\partial M$  of the 4 dimensional space-time manifold $M$ and $P$ is the fluid pressure, given by $P=\alpha \rho$, where $\rho$ is the density of the fluid and $\alpha$ is a constant. The units are so chosen that $16\pi G = 1.$

The Kantowski-Sachs metric is given by 
\begin{equation}
\label{Metric}
ds^{2}=-n^{2}dt^{2}+X^{2}(t)dr^{2}+Y^{2}(t)\left[d\theta^{2}+\sin^{2}\left(\theta\right)d\phi^{2}\right],
\end{equation}
where $n(t)$ is a lapse function while $X(t)$ and $Y(t)$ are scale factors. In order to facilitate the process of quantization, we go over to the Misner representation via the  transformation\cite{Bastos}
\begin{eqnarray}
\label{misner}
X(t)=e^{\sqrt{3}\beta_{+}},\\
Y(t)=e^{-\sqrt{3}\left(\beta_{+}+\beta_{-}\right)}.
\end{eqnarray}.
This transformation allows us to write the metric (\ref{Metric}) as

\begin{equation}
\label{Misner}
ds^{2}=-n^{2}dt^{2}+e^{2\sqrt{3}\beta_{+}}dr^{2}+e^{-2\sqrt{3}\left(\beta_{+}+\beta_{-}\right)}\left[d\theta^{2}+\sin^{2}\left(\theta\right)d\phi^{2}\right].
\end{equation}
The fluid sector can be expressed in terms of some thermodynamic variables following Schutz formalism\cite{schutz1, schutz2}, which has been subsequently developed by Lapchinskii and Rubakov\cite{rubakov} in context of quantum cosmology. The method has been used quite extensively later by many \cite{ sridip1, sridip2, alvarenga1, alvarenga2, barun}. In this formalism, the Lagrangian for the system can be extracted out of the action (\ref{Action}) as 

\begin{equation}
\label{lagrangian}
{\mathcal L}=\frac{6}{n}e^{-\sqrt{3}\left(\beta_{+}+2\beta_{-}\right)}\left[\dot{\beta}_{+}^{2}-\dot{\beta}_{-}^{2}\right]
+ 2n e^{\sqrt{3}\beta_{+}} + \left[n^{-\frac{1}{\alpha}}e^{3\beta_{0}}\frac{\alpha}{\left(1+\alpha\right)^{1+\frac{1}{\alpha}}}\left(\dot{\epsilon}+\theta\dot{S}\right)^{1+\frac{1}{\alpha}}e^{-\frac{S}{\alpha}}\right].
\end{equation}
The metric as well as all other quantities are spatially homogeneous, the integration of space part yields a constant in (\ref{Action}) and is thus inconsequential as it can be absorbed in the right hand side zero of the variational principle. The quantities $\theta$, $\epsilon$, $s$ are thermodynamic potentials which determine the velocity vector. The details are given in \cite{rubakov}. The last term within the square bracket is the contribution from the fluid sector whereas the rest denotes contribution from the gravity sector. The corresponding Hamiltonian for the gravity sector for the metric (\ref{Misner}) can be written as
\begin{equation}
\label{hamiltoniangr}
H_{g}=\frac{n}{24}e^{\sqrt{3}\left(\beta_{+}+2\beta_{-}\right)}\left[-p_{\beta_{-}}^{2}+p_{\beta_{+}}^{2}-48e^{-2\sqrt{3}\beta_{-}}\right],
\end{equation}
where the canonical momenta are defined in the usual way.
We define the canonical momenta for the fluid sector as  $p_{\epsilon}=\frac{\partial{\mathcal L}_{f}}{\partial\dot{\epsilon}}$ and
$p_{S}=\frac{\partial{\mathcal L}_{f}}{\partial\dot{S}}$ and the corresponding Hamiltonian is written as
\begin{equation}
\label{hamfl}
H_{f}=ne^{-3\alpha\beta_{0}}p_{\epsilon}^{\alpha +1}e^{S}.
\end{equation}
With the canonical transformation,
\begin{eqnarray}\label{9}
T&=&-p_{S}\exp(-S)p_{\epsilon}^{-\alpha -1},\\
p_{T}&=&p_{\epsilon}^{\alpha+1}\exp(S),\\
\epsilon^{\prime}&=&\epsilon+\left(\alpha+1\right)\frac{p_{S}}{p_{\epsilon}},\\
p_{\epsilon}^{\prime}&=&p_{\epsilon},
\end{eqnarray}
the Hamiltonian for the fluid sector becomes
\begin{equation}
\label{hamiltonianfl}
H_{f}=ne^{\alpha\sqrt{3}\left(\beta_{+}+2\beta_{-}\right)}p_{T}.
\end{equation}
Hence combining (\ref{hamiltoniangr}) and (\ref{hamiltonianfl}), the super Hamiltonian is given by
\begin{eqnarray}
\label{superhamiltonian}
\nonumber H=H_{g}+H_{f}\\
=\frac{ne^{\alpha\sqrt{3}\left(\beta_{+}+2\beta_{-}\right)}}{24}\left[e^{\sqrt{3}\left(1-\alpha\right)\left(\beta_{+}+2\beta_{-}\right)}\left\{-p_{\beta_{-}}^{2}+p_{\beta_{+}}^{2}-48e^{-2\sqrt{3}\beta_{-}}\right\}+24p_{T}\right]
\end{eqnarray}

Now, one can choose a gauge $n=24n_{0}e^{-\alpha\sqrt{3}\left(\beta_{+}+2\beta_{-}\right)}$ and vary (\ref{superhamiltonian}) with respect to $n_{0}$ to obtain the Hamiltonian constraint as
\begin{equation}
\label{constraint}
e^{\sqrt{3}\left(1-\alpha\right)\left(\beta_{+}+2\beta_{-}\right)}\left\{-p_{\beta_{-}}^{2}+p_{\beta_{+}}^{2}-48e^{-2\sqrt{3}\beta_{-}}\right\}+24p_{T}=0.
\end{equation}

\section{Stiff fluid ($\rho=P$)}

As an example, we choose a stiff fluid given by $\alpha=1$. This choice avoids the ordering ambiguity since $e^{\sqrt{3}\left(1-\alpha\right)\left(\beta_{+}+2\beta_-\right)}=1$ for $\alpha=1$. On quantizing the model, we have the Wheeler-deWitt equation,
\begin{equation}
\label{wd}
\left\{\frac{\partial^{2}}{\partial\beta_{-}^{2}}-\frac{\partial^{2}}{\partial\beta_{+}^{2}}-48e^{-2\sqrt{3}\beta_{-}}\right\}\psi =24\imath\frac{\partial\psi}{\partial T},
\end{equation}
in units of ${\hbar}=1$. The canonical momenta $p_{\beta_i}$ and $p_{T}$ are replaced by $-\imath \frac{\partial}{\partial  \beta_{i}}$ and $-\imath \frac{\partial}{\partial T}$ respectively following the procedure elucidated in \cite{alvarenga3, barun, sridip1}. \\

\subsection{Probability density}

With the separability ansatz $ \psi=e^{\imath \sqrt{3} k_{+}\beta_{+}}\phi(\beta_{-})e^{-\imath E T},$ (\ref{wd}) becomes
\begin{equation}
\label{wd1}
\left\{\frac{\partial^{2}}{\partial\beta_{-}^{2}}+3k_{+}^{2}-48e^{-2\sqrt{3}\beta_{-}}\right\}\phi =24E\phi,
\end{equation}
which has the solution
\begin{equation}
\phi = K_{\imath\nu}\left(4e^{-\sqrt{3}\beta_{-}}\right),
\end{equation}
where $\nu \equiv \sqrt{k_{+}^{2}-8E}$. It deserves mention that this solution is similar to that obtained by Garcia-Compean {\it et al}\cite{ramirez}. The expression for $\nu$ in the present work is different due to the contribution from the fluid sector. \\

Now, we note that $\beta_+$ sector admits a plane wave like solution. Hence, we might be tempted to form a Gaussian wavepacket superposing different plane wave solutions by doing an integration over $k_{+}$ variable. But this is actually not that simple. If we want to construct a wavepacket with definite energy $E$, then varying $k_{+}$ would imply varying $\nu =\sqrt{k_{+}^{2}-8E}$, hence we are forced to superpose $K_{\imath\nu}\left(4e^{-\sqrt{3}\beta_{-}}\right)$ 's as well, resulting to a complicated wavepacket, which is not Gaussian at all. This implies that if we intend to form a Gaussian wavepacket corresponding to $\beta_{+}$ sector, it can not be an energy eigenfunction, which motivates us to define a new parameter
\begin{equation}\label{def}
\sigma \equiv k_{+}^{2}-8E=\nu^{2}
\end{equation}
and trade off $(E,k_+)$ against $(\sigma,k_{+})$ i.e  the wavefunction can be thought of dependent on $k_+$ and $\sigma$ independently, and $E$ becomes a parameter dependent on $k_+$ and $\sigma$ through equation (\ref{def}). \\

In terms of $\sigma$, the wavefunction for $\beta_{-}$ sector reads as
\begin{equation}
\phi = K_{\imath\sqrt{\sigma}}\left(4\chi \right),
\end{equation}
where we have defined, for brevity, $\chi\equiv e^{-\sqrt{3}\beta_{-}}$.\\

Now we can superpose the plane wave like solutions for $\beta_+$ sector , keeping $\sigma$ fixed. This, in turn, means the resulting wavepacket will no more be an energy eigenstate, it is rather a superposition of energy eigenstates. The wavepacket is given by following expression,

\begin{equation}
\Psi = \phi(\chi)e^{\imath\frac{\sigma}{8}T} \psi(\beta_+),
\end{equation} 
where we define $\psi(\beta_+)$ as 
\begin{equation}
\psi(\beta_+)=\int dk_{+} e^{-\left(k_{+}-k_{+0}\right)^{2}}e^{\imath \sqrt{3} k_{+}\beta_{+}-\imath \frac{k_{+}^{2}}{8}T}.
\end{equation}

If we want to insist on superposition of negative energy eigenstates only, the integral over $k_{+}$ should be in the interval $\left(-\sqrt{\sigma},\sqrt{\sigma}\right)$.The requirement of a negative $E$ for the gravity sector stems from the demand that the energy for the fluid part is positive whereas the total energy is zero in view of the Hamiltonian constraint. Similar kind of consideration has been made in earlier work \cite{sridip1, sridip2}. \\

Upon integration over the said interval, we have
\begin{equation}
\psi =\frac{1+\imath}{\sqrt{-8\imath+T}}\sqrt{\pi}\exp\left[\frac{-k_{+0}^{2}T+8\sqrt{3}k_{+0}\beta_{+}+6\imath\beta_{+}^{2}}{-8\imath +T}\right]\left[g(8)-g(-8)\right]
\end{equation}
where $g(a)$ is given by 
\begin{equation}
g(a)=Erfi\left[\frac{\frac{1+\imath}{4}\left(8k_{+0}+\sqrt{\sigma}(-a-\imath T)+4\imath\sqrt{3}\beta_{+}\right)}{\sqrt{-8\imath +T}}\right] 
\end{equation}

With the definition,
$||\psi ||\equiv\int_{-\infty}^{\infty} d\beta_{+}\psi\psi^{*}$, we have 

\begin{eqnarray}
||\psi ||=\frac{1}{\sqrt{3}} \int_{-\sqrt{\sigma}}^{\sqrt{\sigma}} dk_{+} e^{-2\left(k_{+}-k_{+0}\right)^{2}} \nonumber \\
= \frac{1}{2\sqrt{3}}\sqrt{\frac{\pi}{2}}\left(Erf\left[\sqrt{2}(\sqrt{\sigma}+k_{+0})\right]-Erf\left[\sqrt{2}(-\sqrt{\sigma}+k_{+0})\right]\right).
\end{eqnarray}

We have specifically shown, for the  $\beta_+$ sector, that the norm is finite and time independent. This happens non-trivially since we are not dealing with energy eigenstate. We recall that for energy eigenstate, the probability density function is trivially time independent. Now we turn our attention to $\beta_-$ sector. We recast the equation (\ref{wd1}) using the variables $\chi$ and $\sigma$:
\begin{equation}
\chi^2\frac{d^2\phi}{d\chi^2}+\chi\frac{d\phi}{d\chi}+(16\chi^2 - \sigma)\phi = 0
\end{equation}
This equation can be rewritten in the standard self-adjoint form: 
\begin{equation}
\frac{d}{d\chi}\left(\chi\frac{d\phi}{d\chi}\right) +\left(16\chi -\frac{\sigma}{\chi} \right) =0
\end{equation}
with definition of inner product is given by 
\begin{equation}
\langle \phi_{1}| \phi_{2} \rangle \equiv \int_0^{\infty} d\chi \ \chi \  \phi_{1}^{*}(\chi) \phi_2(\chi)
\end{equation}

Hence, the Hamiltonian for $\beta_-$ sector is self-adjoint as well ensuring a unitary time evolution. The norm can explicitly be calculated and shown to be finite, since
\begin{equation}
\int_{0}^{\infty} d\chi\ \chi\ \phi^{*}(\chi) \phi(\chi) =\frac{\pi \nu}{32\sinh\left(\pi \nu\right)}.
\end{equation}
where $\phi(\chi)=K_{\imath\nu}\left(4\chi\right)$ and $\nu^{2}=\sigma$.
\\

The relevant plots of probability density function is given below at different time instant. We note that, $\phi$ is time independent, hence the probability density due to $\beta_{-}$ sector (we will call it $Pr_{-}\equiv \chi \phi(\chi)\phi^{*}(\chi)$) is time independent, while probability density  due to $\beta_{+}$ sector, which we call $Pr_{+}=\psi(\beta_+)\psi^{*}(\beta_+)$, is time dependent (even though the norm is finite and time independent).

  \begin{figure}[!h]
  \begin{subfigure}[b]{0.3\textwidth}
  \centering
 \includegraphics[scale=0.4]{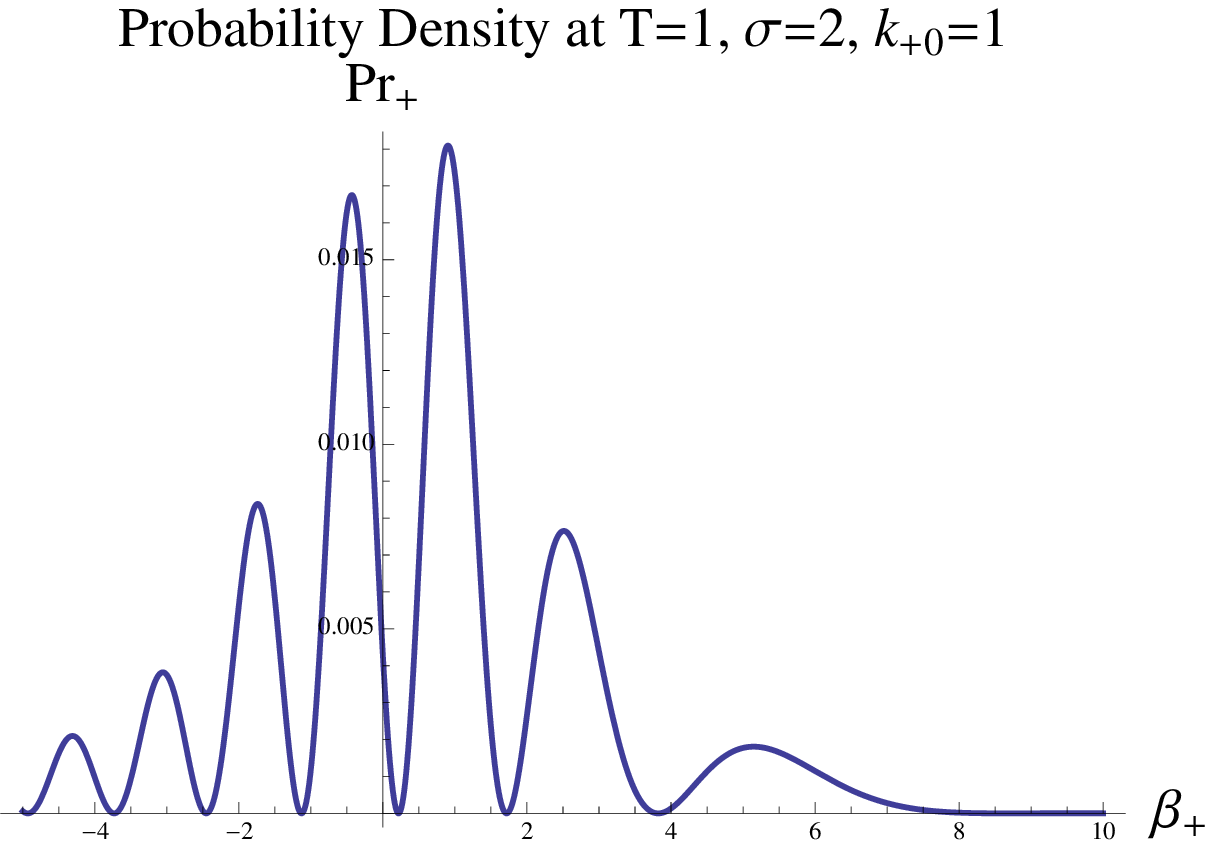}
 \end{subfigure}
   \begin{subfigure}[b]{0.3\textwidth}
  \centering
 \includegraphics[scale=0.4]{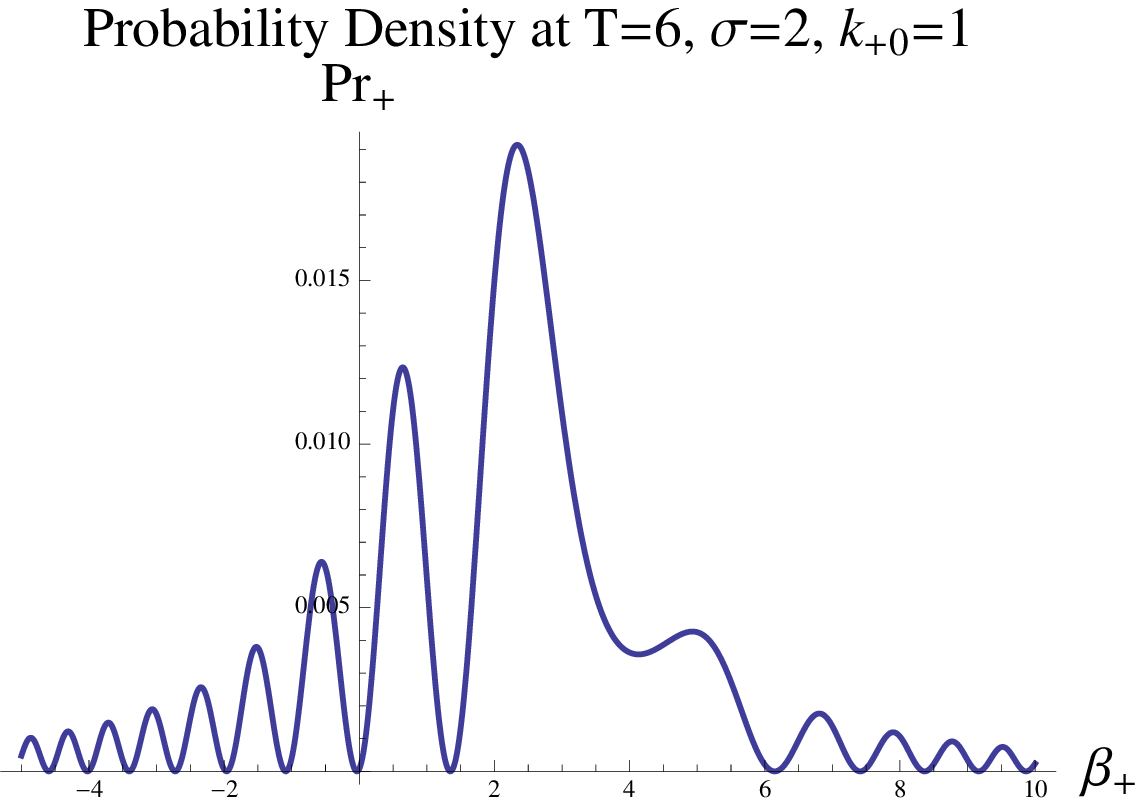}

 \end{subfigure}
   \begin{subfigure}[b]{0.3\textwidth}
  \centering
 \includegraphics[scale=0.4]{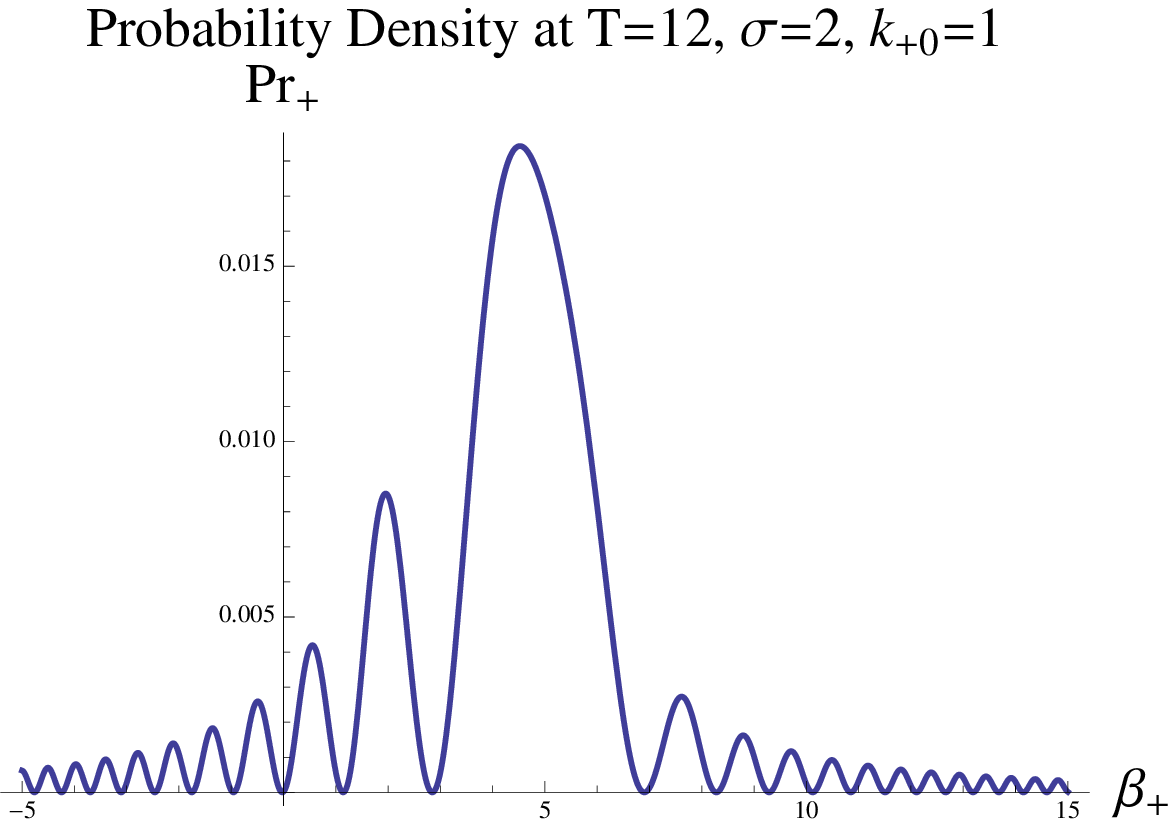}

 \end{subfigure}
  \caption{Probability Density ($Pr_+$)}
\end{figure}
 
    \begin{figure}[!h]
  \centering
 \includegraphics[scale=0.6]{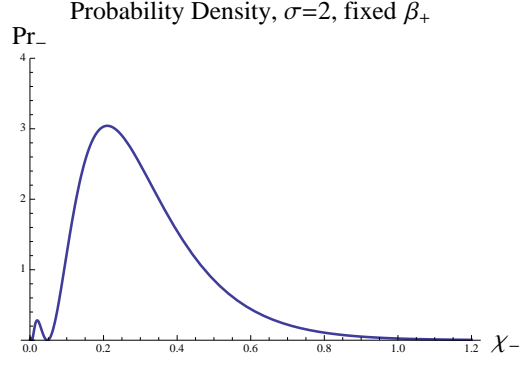}
 \caption{Probability Density ($Pr_-$)}
\end{figure}

We note that the wavefunction dies out for large $\chi$, which translates to the fact that the probability of having $\beta_{-}=-\infty$ is zero. Similarly the probability density dies out as $\beta_{+}$ approaches $-\infty$. This essentially means, the volume of the universe, i.e, $e^{\sqrt{3}\left(\beta_{+}+2\beta_{-}\right)}$ does not hit singularity at any time. Similarly one can argue that the wave function dies out for $\chi=0$ which does imply the probability of large $\beta_{-}$ is very small. But we note, $Pr_{+}$ is time dependent and peak of $Pr_{+}$ does shift to the right with time. Hence, the probability of having large $\beta_{+}$ is not small if we give long enough time for evolution. Hence, the volume can get arbitrarily large with time. 

\subsection{Anisotropy}
The scalar $\tilde{\sigma}^{2}=\frac{1}{2}\sigma^{\mu\nu}\sigma_{\mu\nu}$ found out of shear tensor $\sigma_{\mu\nu}$, can be written for a diagonal metric as,
\begin{eqnarray}
\nonumber \tilde{\sigma}^{2} &=&
\frac{1}{12n^{2}}\left[\left(\frac{\dot{g}_{11}}{g_{11}}-\frac{\dot{g}_{22}}{g_{22}}\right)^{2}
+ \left(\frac{\dot{g}_{22}}{g_{22}}-\frac{\dot{g}_{33}}{g_{33}}\right)^{2}+
\left(\frac{\dot{g}_{33}}{g_{33}}-\frac{\dot{g}_{11}}{g_{11}}\right)^{2}\right]
\end{eqnarray}
which yields
\begin{equation}
\tilde{\sigma}^{2}= \frac{1}{2}e^{-2\sqrt{3}\left(\beta_{+}+2\beta_{-}\right)}\left(2p_{\beta_+}-p_{\beta_-}\right)^{2} \geq 0,
\end{equation}
for the metric (\ref{misner}) where $p_{\beta_\pm}$ are conjugate variables to $\beta_{\pm}$.\\

\section{Models with other Equations of State($0 < \alpha <1 $)}
In this section, we turn our attention to $\alpha \neq 1$. For $\alpha\neq 1$ we rewrite (\ref{constraint}) with the following operator ordering,

\begin{eqnarray}
\label{orderedconstraint}
[e^{\frac{\sqrt{3}}{2}\left(1-\alpha\right)\left(\beta_{+}+4\beta_{-}\right)}p_{\beta_{+}}e^{\frac{\sqrt{3}}{2}\left(1-\alpha\right)\beta_{+}}p_{\beta_{+}}-e^{\sqrt{3}\left(1-\alpha\right)\left(\beta_{+}+\beta_{-}\right)}p_{\beta_{-}}e^{\sqrt{3}\left(1-\alpha\right)\beta_{-}}p_{\beta_{-}} \nonumber \\
 -48e^{-2\sqrt{3}\beta_{-}}e^{\sqrt{3}\left(1-\alpha\right)\left(\beta_{+}+2\beta_{-}\right)}+24p_{T}]=0.
\end{eqnarray}

The guiding principle behind choosing such operator ordering is to ascertain self-adjoint extension which is explained later in this section.

By replacing the momenta by the corresponding operators as described before, we have the Wheeler-DeWitt equation as
\begin{eqnarray}
\label{wd3}
[-e^{\frac{\sqrt{3}}{2}\left(1-\alpha\right)\left(\beta_{+}+4\beta_{-}\right)}\frac{\partial}{\partial \beta_{+}}e^{\frac{\sqrt{3}}{2}\left(1-\alpha\right)\beta_{+}}\frac{\partial}{\partial \beta_{+}}+e^{\sqrt{3}\left(1-\alpha\right)\left(\beta_{+}+\beta_{-}\right)}\frac{\partial}{\partial \beta_{-}}e^{\sqrt{3}\left(1-\alpha\right)\beta_{-}}\frac{\partial}{\partial \beta_{+}} \nonumber \\  -48e^{-2\sqrt{3}\beta_{-}}e^{\sqrt{3}\left(1-\alpha\right)\left(\beta_{+}+2\beta_{-}\right)}]\Psi =24\imath\frac{\partial\Psi}{\partial T}.
\end{eqnarray}

We now effect a transformation of variables as
\begin{eqnarray}
\chi_{+}\equiv e^{-\frac{\sqrt{3}}{2}\left(1-\alpha\right)\beta_{+}},\\
\chi_{-}\equiv e^{-\sqrt{3}\left(1-\alpha\right)\beta_{-}}
\end{eqnarray}
and use separability ansatz $\Psi= \phi(\chi_{+},\chi_{-}) e^{-\imath E T}$ to obtain

\begin{equation}\label{wd5}
H_{g}\phi =-\frac{1}{\chi_{-}^{2}}\frac{\partial^{2}\phi}{\partial\chi_{+}^{2}}+\frac{1}{\chi_{+}^{2}}\frac{\partial^{2}\phi}{\partial\chi_{-}^{2}}-48\chi_{-}^{\frac{2\alpha}{1-\alpha}}\chi_{+}^{-2}\phi = 24E\phi.
\end{equation}

This equation is not apparently separable in $\chi_{\pm}$ and thus one cannot investigate the behaviour analytically. But under a reasonable assumption, we can show the Hamiltonian given by (\ref{wd5}) actually admits a self-adjoint extension. The assumption is that the Hamiltonian is a symmetric operator, by which we mean that the solution  (if it admits any) obeys following conditions 

\begin{equation}
\label{assum}
\left[\phi\frac{\partial\phi^{*}}{\partial \chi_{\pm}}-\phi^{*}\frac{\partial\phi}{\partial \chi_{\pm}}\right]_{0}^{\infty}=0
\end{equation}

To facilitate a clear view, we can think of this assumptions in following fashion. Imagine, we obtain the solution to eq. (\ref{wd5}) and restrict the solution space so as to satisfy eq. (\ref{assum}). Now what we call a reasonable assumption is that this restricted solution space is not null. This is actually quite natural in context of quantum mechanics, we assume similar conditions while we work on a free particle on a half-line or the whole real line. In case of whole line, $0$ in the lower limit of the equation (\ref{assum}) is replaced by $-\infty$. The condition is reasonable in the present context, since, for example, any normalizable solution which vanishes at $\beta_{\pm}=\pm \infty$ will satisfy the condition.

Now we can use Von Neumann's theorem \cite{reed} asserting that a symmetric operator $\hat{A}$ defined on domain $\mathcal{D}$  has equal deficiency index, if there exists a norm preserving anti-unitary conjugation map $C:\mathcal{D}\rightarrow\mathcal{D}$ such that $[\hat{A},C]=0$, which, in turn, shows that $A$ admits self-adjoint extension.\\

It is easy to see that all the conditions for employing Neumann's theorem are satisfied:
\begin{enumerate}
\item Complex conjugation map, $C:\mathcal{Hi}\rightarrow\mathcal{Hi}$, between Hilbert space $\mathcal{Hi}$, takes $\phi$ to $\phi^{*}$, which also belongs to the Hilbert space $\mathcal{Hi}$.\\
\item $C$ is norm preserving , since whatever be the definition of norm, it involves $\phi\phi^{*}$, hence does not change under $C$.
\item $C$ is anti-unitary and we have $CH_{g}=H_{g}C$.
\end{enumerate}

Hence, the theorem goes through and we have self-adjoint extension and thereby a unitary evolution for Kantowski-Sachs cosmology. The theorem can be understood in following manner as well. Since, $\phi_{+}$ satisfies,
\begin{equation}
H_{g}\phi_{+}=\imath\phi_+
\end{equation}
if and only if
\begin{equation}
H_{g}\phi_{-}=-\imath\phi_-
\end{equation}
where $\phi_-=\phi_{+}^{*}=C\phi_+$; the map $C$ induces a one-to-one map between two spaces, whose dimensions are actually named {\bf Deficiency Index} and thereby make them equal, and we know that if a symmetric operator has equal deficiency indices, it does admit a self-adjoint extension. The detailed and rigorous proof can be found in \cite{reed}. It deserves mentioning that the extension may modify the boundary condition (\ref{assum}) by making it more strict (i.e the modified condition will imply eq. (\ref{assum}), not necessarily implied by eq. (\ref{assum}).)\\

The role of operator ordering is crucial in the sense that with such an ordering we have the kinetic term $\frac{\partial^{2}\phi}{\partial \chi_{\pm}^{2}}$ multiplied with $\chi_{\mp}^{2}$. Hence, the condition (\ref{assum}) for its being symmetric is same as the condition for a standard Laplacian since the derivative with respect to $\chi_+$ term is multiplied with $\chi_-$ and vice versa. \\

\section {Discussion}
It has been shown that like anisotropic cosmological models with constant spatial curvature such as Bianchi I, V and IX, the Kantowski-Sachs cosmology can also be quantized where the problem of non-unitarity can be successfully eradicated. The trick is to figure out the correct weight factor while defining the norm and the inner product. This has been explicitly demonstrated for a stiff fluid given by $P=\rho$, i.e., for $\alpha=1$. We have calculated the shear scalar $\tilde{\sigma}^{2}$ which is a positive definite quantity indicating that this unitarity restoration is not at the cost of anisotropy itself. In the case of a Bianchi I quantum cosmology, it had already been shown that the unitarity is not purchased at the expense of anisotropy\cite{sridip3}. For $\alpha\neq1$, however, we could not solve the equation due to inseparability but we have shown that the Hamiltonian admits a self-adjoint extension  (with a suitable operator ordering). Thus the model can actually have a unitary evolution for any ideal barotropic ($P=\alpha \rho$) fluid. \\

One explicit example, namely that with a stiff fluid, along with the implicit proof for non-stiff fluids is good enough to disprove the folklore that anisotropic quantum cosmology necessarily involves non-unitarity. Here we have shown the Hamiltonian is indeed self-adjoint for $\alpha=1$, thereby has to admit a unitary evolution. Furthermore, we have constructed explicit wave-packet showing unitary evolution explicitly. This restoration of unitarity is non-trivial in nature since the wave packet leads to a time dependent probability density, which integrates to a time independent finite quantity, identified as norm, leading to a conservation of probability. This scenario is in complete contrast with having energy eigenstates which trivially yields time independent probability density and thereby time independent norm irrespective of self-adjointness of underlying Hamiltonian. Energy eigenstates, thus, may not actually serve as a signature of unitaritity. \\

For $\alpha\neq 1$, the proof is implicit in nature, it does not shed light on how to obtain a solution or how to do a self-adjoint extension. Yet it is good enough to consolidate the idea that non-unitarity is not generic to anisotropic models, it can be cured with suitable operator ordering or introducing a correct weight factor. \\

As a bonus, it is also interesting to note that the stiff fluid model, a bonafide model which does not violate the classical energy condition, can actually give rise to a singularity free cosmology at the quantum level. This stems from the fact that scale factors have some intrinsic quantum fluctuation around its average value and the wavefunction goes to zero whenever scale-factor hits $0$.\\


\end{document}